\documentclass[aps,prb,twocolumn,showpacs,amsmath,amssymb,floatfix,superscriptaddress]{revtex4}
\usepackage{epsfig}
\usepackage{bm}
\usepackage{color}

\begin{document}

\title{Relaxation of an electron wave packet at the quantum Hall edge at filling factor $\nu=2$ }

\author{Artur O. Slobodeniuk}
\affiliation{D\'epartement de Physique Th\'eorique, Universit\'e de Gen\`eve, CH-1211 Gen\`eve 4, Switzerland}
\affiliation{Laboratoire National des Champs Magn\'etiques Intenses, CNRS-UJF-UPS-INSA, 25, rue des Martyrs, 
             B.P. 166, Grenoble 38042, France}

\author{Edvin G. Idrisov}
\affiliation{D\'epartement de Physique Th\'eorique, Universit\'e de Gen\`eve, CH-1211 Gen\`eve 4, Switzerland}

\author{Eugene V. Sukhorukov}
\affiliation{D\'epartement de Physique Th\'eorique, Universit\'e de Gen\`eve, CH-1211 Gen\`eve 4, Switzerland}
\date{\today}

\begin{abstract}
In this work, we address the recent experiment [S.\ Tewari {\em et al.}, arXiv:1503.05057v1], where the suppression of phase coherence of a single-electron wave packet created at the edge of a quantum Hall (QH) system at filling factor 2 has been investigated with the help of an electronic Mach-Zehnder (MZ) interferometer. The authors of the experiment have observed an unexpected behavior of phase coherence, that saturates at high energies instead of vanishing,
presumably suggesting the relaxation of a wave packet to the ground state before it arrives to the MZ interferometer. Here, we theoretically investigate this situation using the model of edge states [I. P. Levkivskyi, E.~V.~Sukhorukov, Phys.~Rev.~B 78, 045322 (2008)], which accounts for the strong Coulomb interaction between the two electron channels at the edge of a QH system.
We conclude that the observed phenomenon cannot be explained within this model for the reason that under an assumption of linearity of the electron spectrum at low energies the system remains integrable in terms of the collective charge excitations, and therefore full relaxation to the ground state is not possible, despite strong interactions. As a result, the degree of the phase coherence decreases with energy of the initial state in a power-law manner. Since this does not happen in the experiment, a new physical phenomenon may take place at the edge of a QH state, which deserves further investigations. We support our findings by calculating the energy distribution and the Wigner function of the outgoing non-equilibrium state of the single-electron wave packet.
\end{abstract}

\maketitle

\section{Introduction}
Quantum Hall edge states at integer filling factors \cite{Halperin,Buttiker}  present a notable example of a strongly interacting quasi-one-dimensional system. Typically, in such systems interactions cannot be considered perturbatively, and the associated physical phenomena do not have any analogues in weakly interacting systems. For example,  at filling factor $\nu=2$ the Coulomb interaction between the two copropagating chiral electron channels at the edge of a QH system splits the spectrum of the collective excitations into one dipole and one charged mode, that propagate with different velocities. \cite{Bocquillon_1} As a result, at relatively long length scales, where these two excitations are well separated in space, this leads to strong correlations between electrons. New interesting mesoscopic phenomena that arise from this correlations have recently been the subject of experimental studies and intensive theoretical discussions. In the context of the present paper we would like to mention the observation of the lobes \cite{Neder_1,Litvin_1, Preden, Neder_2, Litvin_2,Bieri, Huynh} and of the shot noise-induced phase transition \cite{Helzel} in the visibility of Aharonov-Bohm (AB) oscillations in electronic MZ interferometers,  and the experimental studies of relaxation of non-equilibrium electron distribution functions. \cite{Degiovanni,Altimiras_1,Sueur, Altimiras_2}

\begin{figure}[]
\includegraphics[width=8cm]{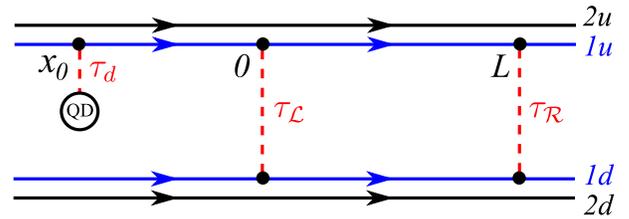}
\caption{ The MZ interferometer, shown in the figure, consists of two edge channels of a QH system at filling factor $\nu=2$, connected
by two tunnel junctions with coupling amplitudes $\tau_\mathcal{L}$ and $\tau_\mathcal{R}$.
A quantum dot with charging energy $\varepsilon_0$ is attached to one of the channels of the QH edge
by the tunnel junction with the coupling amplitude $\tau_d$.}
\vspace{-5mm}
\label{inter}
\end{figure}

The essential ingredient of all mentioned above experiments is that a strongly non-equilibrium state is created by applying a voltage bias to a quantum point contact (QPC) and injecting a current of electrons into one of the two electron channels. The use of this technique has a crucial effect on the subsequent evolution of the propagating state, most prominent example being the mentioned above phase transition when the QPC's transparency is equal to $1/2$. Very recently, the group of Patrice Roche has implemented an alternative approach, \cite{Tewari} where a single-electron wave packet is injected into one of the edge channels by using resonant tunnelling via a single energy level in a quantum dot (QD) (see Fig.\ \ref{inter}). The purpose of this approach was to reduce the effects of the partition noise of the first QPC, and of the energy averaging in the initial state due to the applied voltage bias, and thus to focus solely on the effects of interaction in the created state. These effects have been probed with the help of the MZ interferometer attached to the system downstream of the injection point, as schematically shown in Fig.\ \ref{inter}. One of the puzzling results of this experiment, outlined below, has motivated our present work.

The experimentalists have investigated the dependence of the visibility of the AB oscillation in the MZ interferometer on the energy of the injected electron wave packet. Typically, the phase coherence reduces gradually at high energies, because this increases the possibility for electrons to scatter inelastically. Surprisingly, the paper [\onlinecite{Tewari}] has reported an initial suppression of the visibility starting from zero energy, which is, however, followed by its saturation at the value 0.12 at energy around 25 $\mu eV$ and by a long plateau up to energies of 120 $\mu eV$. The authors have argued that a possible explanation of this unexpected behaviour should account for the fact that strong interactions lead to the relaxation of energy of the initial electron state. When such a state arrives to the interferometer, regardless of the energy of the initial state, it cannot scatter inelastically anymore, which may explain the plateau in the visibility of the AB oscillations.

One of the goals of the present paper is to investigate whether the simple model of strongly interacting QH edge states at filling factor $\nu=2$, proposed earlier in Ref.\ [\onlinecite{Levkivskyi_1}] and proved to be essential to explain the recent experiments mentioned above, is also capable of explaining the experiment in Ref.\ [\onlinecite{Tewari}]. According to this model,  the chiral character of the electron transport at the QH edge allows one to apply the free-fermionic approach to the derivation of the initial state of the injected electron wave packet, followed by the use of the bosonization technique \cite{Giamarchi} to describe its subsequent evolution, and to account for the interactions non-perturbatively. We examine the splitting of the state at the QPCs of the MZ interferometer and evaluate the average current and the visibility of the AB oscillations using the tunnelling Hamiltonian approach to the lowest order in tunnelling. Finally, we come to the conclusion that by using this approach, one is not able to explain the plateau in the visibility, observed in the experiment [\onlinecite{Tewari}], i.e., some additional physical ingredients have to be considered, such as, perhaps,  effects of disorder and/or of the non-linear dispersion, that are not included in our model. Instead, we find that the visibility as a function of the energy of the initial state, oscillates and decays in a power-law manner.

Our results can be interpreted in two alternative ways. On one hand, despite the strong interactions, a chiral quasi-one-dimensional system of electrons remains an integrable system in terms of collective bosonic charge excitations. In other words, in the case of linear dispersion law for electrons, the bosons are free and, therefore, they keep the memory of the initial state, including its energy. Thus the plateau in the visibility of the AB oscillations as a function of the energy of the initial state is not expected. On the other hand, although the initial electron wave packet splits into two wave packets propagating with different velocities, each of them does not spread further due to the linear dispersion law, as we show below. Therefore, all the initial energy is stored in the space limited by the initial width of the wave packets, and mostly in the form of electron-hole excitations. This heats the state up to the temperatures of the order of the initial energy, and leads to energy dependent decoherence.

Our findings with regards to the important role played by the integrability raise an interesting question about the nature of the intermediate quasi-stationary state, i.e., before it reaches the final equilibrium. Indeed, if the initial electron wave packet does not relax to the Fermi level despite the strong interactions, what is the energy distribution in the outgoing state? In order to investigate this question in detail, we follow the suggestion of the Refs.\~[\onlinecite{Ferraro_1,Ferraro_2}] and derive the energy distribution and the Wigner function of the two outgoing wave packets corresponding to the dipole and the charged mode. We find that the effect of interactions is twofold: It distributes an electron with an initial energy $\varepsilon_0$ more or less homogeneously in the interval between $-\varepsilon_0$ and $\varepsilon_0$  around the Fermi level, and it creates a bunch of electron-hole excitations with the excess distribution $f(\varepsilon)-\theta(-\varepsilon) \propto 1/\varepsilon$ with a cutoff at $|\varepsilon |=\varepsilon_0$. These our findings may be of great importance for the new field of electron quantum optics, \cite{Henny, Neder_3,Ji,Roulleau_1,Roulleau_2,Bocquillon_3} which is essentially based on creating electron states with the help of single-electron sources. \cite{Geerligs,Pothier,Kouwenhoven,Fujiwara,Feve,Bocquillon_2}

It is worth to mention that a similar problem has been recently addressed \cite{Levkivskyi_2,Levkivskyi_3} in the context of the experiment of the group of Frederic Pierre, \cite{Degiovanni,Altimiras_1,Sueur, Altimiras_2} where the initial non-equilibrium state was created at the QH edge by injecting electrons via a QPC of transparency $\cal T$, biased with a voltage $V$.  The Ref.\ [\onlinecite{Levkivskyi_3}] finds that in the case of low transparencies, ${\cal T} \ll 1$,  the initial double-step electron distribution function relaxes to the quasi-stationary distribution of the form $f(\varepsilon)-\theta(-\varepsilon)\propto 1/\varepsilon$ with a high-energy cutoff at $|\varepsilon|=eV$, and a low-energy cutoff at $|\varepsilon|\sim eV{\cal T}$. Taking into account our present results, this behaviour can be interpreted as a result of the individual contributions of electron wave packets injected by the QPC, and of the collective effect of the partition noise of the QPC, regularizing the distribution at low energies.

The rest of the paper is organised as follows. In Sec.\ \ref{Sec:I}  we introduce the model of the system, starting from the Hamiltonian of all the constituting parts, followed by the bosonization prescription.  In Sec.\ \ref{Sec:II} we derive the initial state of the electron wave packet injected into one of the edge channels. In Sec.\ \ref{Sec:III} we formulate the tunnelling Hamiltonian method and derive the direct and the interference terms in the average current through the MZ interferometer, as well as the visibility of the AB oscillations. Sec.\ \ref{Sec:VII} is devoted to the derivation of the Wigner function and of the distribution function of the outgoing state. We present our conclusions in Sec.\ \ref{Sec:VIII}.

\section{Model}
\label{Sec:I}

We start by introducing the Hamiltonian of a QD connected to an interferometer (see Fig.\,\ref{inter}).
We use the effective theory describing edge states as collective fluctuations of the charge density
$\rho_{\alpha j}(x)$, where indexes $\alpha=1,2$ label the number of the channel in the state $j=u,d$
\begin{equation}
\label{Hamiltonian}
H=H_0+H_d+H_{tun,d}+H_{tun}.
\end{equation}
Here
\begin{equation}
H_0=\frac{1}{2}\int dx \sum_{\alpha\beta,j} V_{\alpha\beta}\rho_{\alpha j}(x)\rho_{\beta j}(x)
\end{equation}
is the Hamiltonian of the quantum Hall edge states at filling factor $\nu=2$.
The charge density operators are expressed in terms of chiral bosonic fields $\phi_{\alpha j}(x)$, namely,
$\rho_{\alpha j}(x)=(1/2\pi)\partial_x\phi_{\alpha j}(x)$. These fields satisfy commutation relations
\begin{equation}
\label{comm}
  [\phi_{\alpha j}(x),\phi_{\beta k}(y)]=i\pi\delta_{\alpha\beta}\delta_{jk}sgn(x-y).
 \end{equation}
The electron operator in each channel can be presented as
\begin{equation}
\label{fermion-boson-correspondance}
\psi_{\alpha j}(x)=\frac{1}{\sqrt{a}}e^{i\phi_{\alpha j}(x)},
\end{equation}
where $a$ is an ultraviolet cutoff. Eq.\ (\ref{comm}) guarantees the fermionic commutation relations of the operators (\ref{fermion-boson-correspondance}), and that these operators add and remove a charge equal to 1 (in units of $e$).

 The matrix $V_{\alpha\beta}$ has the form \cite{Levkivskyi_3}
  \begin{equation}
 V=\left(\begin{array}{cc}
   U+2\pi v & U \\
   U & U+2\pi v
   \end{array}
\right).
\end{equation}
were $U>0$ defines the strong screened Coulomb interaction, and $v$ is the velocity of the edge excitations without interaction.
Hamiltonian $H_0$ can be diagonalized by the transformation
\begin{equation}
 \phi_{1j}=\frac{1}{\sqrt{2}}(\xi_{1j}+\xi_{2j}),\quad \phi_{2j}=\frac{1}{\sqrt{2}}(\xi_{1j}-\xi_{2j}),
\end{equation}
introducing the charged $\xi_{1j}(x)$ and the dipole $\xi_{2j}(x)$ modes, respectively.
The new fields satisfy the commutation relations
\begin{equation}
  [\xi_{\alpha j}(x),\xi_{\beta k}(y)]=i\pi\delta_{\alpha\beta}\delta_{jk}sgn(x-y).
 \end{equation}
 Substituting these fields into the Hamiltonian, we obtain
   \begin{equation}
   H_0=\frac{1}{4\pi}\int dx \sum_{\alpha,j} v_\alpha (\partial_x\xi_{\alpha j})^2,
  \end{equation}
   where $v_1=u=U/\pi+v$ and $v_2=v$.

   The Hamiltonian of the QD at the resonance with the Fermi level may be written as:
   \begin{equation}
   H_{d}=\varepsilon_0d^\dag d,
   \end{equation}
   where $\varepsilon_0$ is the energy of the charged QD, with respect to the Fermi level in the edge channels,
   and $d$ is the electron annihilation operator at the dot.
   The Hamiltonian of tunneling between the dot and the channel $1u$ reads
   \begin{equation}
   \label{H_tun_d}
 H_{tun,d}=\tau_d\psi_{1u}^\dag(x_0)d+\mathrm{H.c.},
\end{equation}
where $\tau_d$ is the coupling constant.
Using Eq.\ (\ref{fermion-boson-correspondance}), it can be written as
 \begin{equation}
 H_{tun,d}=\frac{\tau_d}{\sqrt{a}}e^{-i\phi_{1u}(x_0)}d+\mathrm{H.c.},
\end{equation}

  Finally, the last term of Eq.\ (\ref{Hamiltonian}),
  \begin{equation}
  H_{tun}= \sum_{\ell=\mathcal{L,R}} A_\ell +\mathrm{H.c.},
\end{equation}
describes the tunneling between the edge channels $1u$ and $1d$. Here
 \begin{equation}
A_\ell=\tau_\ell\psi_{1u}^\dag(x_\ell)\psi_{1d}(x_\ell),
\end{equation}
and $x_\mathcal{L}=0$ and $x_\mathcal{R}=L$, i.e., the tunnel junctions are placed at the points $x=0$ and $x=L$.
The bosonic representation of the tunneling Hamiltonian then acquires the form
\begin{equation}
   H_{tun}=\sum_{\ell=\mathcal{L,R}}\frac{\tau_\ell}{a}\, e^{-i\phi_{1u}(x_\ell)}e^{i\phi_{1d}(x_\ell)}+\mathrm{H.c.}.
\end{equation}

This model describes the evolution of the initial state of the charged QD with energy $\varepsilon_0$.
Analysis of the evolution of this state is a complex problem, because the Coulomb interaction makes the total Hamiltonian non-linear in terms of fermions. We thus split this problem in two parts. First, we consider the process of discharging the QD into the QH edge and obtain the electron state on the edge channel as a function of time $t>0$. Then, we construct an auxiliary initial state in the system  without a QD, but which gives the same state as in the previous case after being evolved by the edge Hamiltonian at the time $t$, and consider it as the initial state for the interferometer. This procedure allows us to consider the first part of the problem non-perturbatively with respect to the coupling constant $\tau_d$. The second part of the problem is then solved as a perturbation theory with respect to the coupling constants $\tau_\mathcal{L}$ and $\tau_\mathcal{R}$. The simplification, justifying this procedure and leading to an exact solution of the problem, arises from the fact that the processes of discharging of the QD and of tunnelling of the excitations in the interferometer take place sequentially, because of the chirality of the edge excitations and of the local nature of the tunneling processes.

\section{Initial state}
\label{Sec:II}
As an intermediate step, we consider the case of filling factor $\nu=1$ and describe the QH edge states using the free-fermion picture.
We obtain the initial state for the interferometer in this case, and then use it to derive the initial state in the case of $\nu=2$.

The total Hamiltonian of a QD, tunnel-coupled to a QH edge,
reads $H_{tot}=H_F+H_{d}+H_{tun,d}$, where
\begin{equation}
 H_F+H_{d}=-iv_F\int dx\, \psi^\dag(x)\partial_x\psi(x)+\varepsilon_0d^\dag d,
\end{equation}
is the Hamiltonian of the QH edge and the QD, and
\begin{equation}
 H_{tun,d}=\tau_d\psi^\dag(x_0)d+\mathrm{H.c.}
\end{equation}
is the tunneling Hamiltonian, $v_F$ is the Fermi velocity.
Let us consider the evolution of the single-particle state
\begin{equation}
 |\Psi(t)\rangle=\int dx f(x,t)\psi^\dag(x)|0\rangle+ C(t)d^\dag|0\rangle,
\end{equation}
where $\psi(x)|0\rangle=d|0\rangle=0$.
Solving the equations of motion
\begin{equation}
 \partial_tf(x,t)=-v_F\partial_x f(x,t)-i\tau_dC(t)\delta(x-x_0),
 \end{equation}
 \begin{equation}
 \partial_t C(t)=-i\varepsilon_0C(t)-i\tau_d^*f(x_0,t),
\end{equation}
with the initial conditions $f(x,0)=0$ and $C(0)=1$, implying that at time $t=0$ the QD is charged
and the channel is empty, we obtain
\begin{align}
C(t)=&e^{-i(\varepsilon_0-i\Gamma)t},\\
\label{Coeffitient}
f(x,t)=&\frac{\tau_d}{iv_F}\theta(x-x_0)\,\theta\!\left(t-\frac{x-x_0}{v_F}\right)e^{-i(\varepsilon_0-i\Gamma)\left(t-\frac{x-x_0}{v_F}\right)},
\end{align}
where $\Gamma=|\tau_d|^2/2v_F$ is the QD level width.
Considering the long-time limit $t=T\gg\Gamma^{-1}$, we approximate the single-particle state
(with exponential accuracy) as
\begin{equation}
 |\Psi(T)\rangle=\int dx f(x,T)\psi^\dag(x)|0\rangle,
\end{equation}
because the QD is empty on large times.

Let us apply the operator $e^{iH_FT}$, evolving the state backward in time:
 \begin{equation}
|\Psi\rangle_{in}=e^{iH_FT}|\Psi(T)\rangle=\int dx f(x+v_FT,T)\psi^\dag(x)|0\rangle.
\end{equation}
It is important that the time evolution of
the initial state $|\Psi\rangle_{in}$ with Hamiltonian $H_F$ therefore results in the single-particle state of the total system
containing the QD and the edge channel at time $t=T\gg\Gamma^{-1}$.
Moreover, one can show that this result is correct for all times $T>0$, because of chirality of the edge excitations.
One can therefore replace the problem of finding the evolution of the electron state injected from the QD,
by the calculation of the evolution of an auxiliary initial state without the QD.
The many-particle case, then can be obtained with the initial state
  \begin{equation}
|\Psi\rangle_{in}^F=\int dx f(x+v_FT,T)\psi^\dag(x)|FS\rangle ,
\end{equation}
where $|FS\rangle$ is the Fermi sea ground state of the system.
Particularly, a state with an electron injected at point $x_0$ then can be presented as
\begin{equation}
\label{Free-fermion-state}
|\Psi\rangle_{in}^F=\frac{|\tau_d|}{v_F}\int_{x_0-v_F T}^{x_0}dxe^{i(\varepsilon_0-i\Gamma)(x-x_0)/v_F}\psi^\dag(x)|FS\rangle,
\end{equation}
where we have omitted an unimportant phase factor. This result is valid for $\varepsilon_0/\Gamma\gg1$.

 The initial state has been obtained for the case of free fermions, therefore it cannot describe the case of two quantum channels with the Coulomb interaction. As a simple manifestation of this difference we can point out that the initial state (\ref{Free-fermion-state}) of the system contains the Fermi velocity $v_F$ of the free fermion problem. However, it should be characterized by parameters of the real channel, such as velocities $u$ and $v$.

  The processes of discharging of the QD and further evolution of the edge excitations in the channel have three important properties. First,  during the tunneling process, the electron excitation appears in the electron channel at the point $x_0$ because of the locality of the tunneling Hamiltonian (\ref{H_tun_d}). Second, the charge in the QD decays exponentially, as in the free-fermion case, but with a different rate $\Gamma$, which is determined by the density of electronic states in the channel. This means that the current in the point $x_0$ also has an exponential profile. Third, the absence of backscattering in the channel and assumed in our model short-range character of the interaction implies the absence of back-action on the QD. Since apart from the renormalization of the parameter $\Gamma$ the tunneling process is not affected by interactions, we find the resulting quantum state using the following method.

We consider a free fermion wave-packet state (\ref{Free-fermion-state}) with certain $\Gamma$ which is found later, and let it evolve during time $T$ with the diagonalized Hamiltonian
\begin{align}
\label{H_1}
H_1=&\frac{1}{4\pi}\int_{-\infty}^{x_0} dx v_F[(\partial_x\xi_1)^2+(\partial_x\xi_2)^2]\nonumber\\ +
&\frac{1}{4\pi}\int_{x_0}^\infty dx [u(\partial_x\xi_1)^2+v(\partial_x\xi_2)^2]
\end{align}
of the system where the interactions are only present from $x=x_0$ to $x\rightarrow\infty$. A state, obtained in a such way is equivalent to the state created in the interacting channel as a result of discharging of the QD. Then, we  apply the backward evolution during time interval $T$ with the interacting Hamiltonian in the diagonalized form
\begin{equation}
\label{H_2}
H_2=\frac{1}{4\pi}\int dx [u(\partial_x\xi_1)^2+v(\partial_x\xi_2)^2]
\end{equation}
to find the initial state in the interacting channel.
Substituting the free-fermion wave packet (\ref{Coeffitient}), we obtain
\begin{align}
\label{Psi_in^B}
|\Psi\rangle_{in}^B=\frac{|\tau_d|}{\sqrt{a}}\int_{-\infty}^0
d\tau e^{i(\varepsilon_0-i\Gamma)\tau}e^{-\frac{i}{\sqrt{2}}\sum_{\alpha}\xi_\alpha(x_0+v_\alpha\tau)}|\Omega\rangle
\end{align}
(see Appendix \ref{Sec:A} for details of the calculation).
Here $|\Omega\rangle=|0\rangle_1|0\rangle_2$ is the ground state of the two channels of the QH edge.

The parameter $\Gamma$ can be found from the normalization condition by pointing out that the appropriate state described by Eq.\ (\ref{Psi_in^B}) is asymptotically equal to the exact state in the limit of $\varepsilon_0/\Gamma\gg1$. The normalization condition then gives
$\Gamma=|\tau_d|^2/2\sqrt{uv}$. We will use this relation in the calculations below.
If the QH edge channel is an element of a larger system (such as an interferometer in our case), we replace the ground state $|\Omega\rangle$ with the ground state of the total system $|\Phi\rangle$, as described in the next section.

\section{Transport through the interferometer}
\label{Sec:III}
In this section we investigate the current through the
interferometer and focus on the behavior of the visibility of the AB oscillations as a function of the parameters of the interferometer
and of the energy of the injected state.
We consider the time evolution of the initial state with the Hamiltonian of the interferometer
\begin{equation}
H_{MZ}=H_0+H_{tun},
\end{equation}
where $H_0$ and $H_{tun}$ are introduced in Sec.\ \ref{Sec:I}, and
consider the average current and the total transmitted charge in the channel $1d$ after the second tunnel junction,
and the dependence of the visibility on the energy $\varepsilon_0$.
The tunneling current operator
\begin{equation}
I(t)=ie^{iH_{MZ}t}[H_{tun},N_{1d}]e^{-iH_{MZ}t},
\end{equation}
is nothing but the rate of change $I(t)=\dot{N}_{1d}(t)$ of the number of electrons
in the edge channel $1d$
\begin{equation}
N_{1d}=\int dx\, \psi^{\dag}_{1d}(x)\psi_{1d}(x).
\end{equation}

As a first step, we obtain the expression for the time-dependent average current
$\mathrm{I}(t)=\langle\Psi|I(t)|\Psi\rangle$, where the average is taken on the initial state
\begin{equation}
|\Psi\rangle=\frac{|\tau_d|}{\sqrt{a}}\int_{-\infty}^0\!\!\! d\tau e^{i(\varepsilon_0-i\Gamma)\tau}e^{{-\frac{i}{\sqrt{2}}}\sum_{\alpha}\xi_{\alpha u}(x_0+v_\alpha\tau)}|\Phi\rangle,
\end{equation}
where $|\Phi\rangle$ is the ground state of the interferometer
\begin{equation}
|\Phi\rangle=U(0,-\infty)|\Omega\rangle_u|\Omega\rangle_d,
\end{equation}
obtained perturbatively by adiabatically applying the tunneling
perturbation to the ground state of disconnected channels. To the lowest order in tunneling,
 \begin{equation}
 U(t,t')=e^{-iH_0(t-t')}\left[1-i\int_{t'}^t dt^{\prime\prime}\widetilde{H}_{tun}(t^{\prime\prime}-t^{\prime})\right].
\end{equation}
The tilde in $\widetilde{H}_{tun}$ denotes the interaction representation operator
\begin{equation}
\widetilde{O}(t)=e^{iH_0t}O(0)e^{-iH_0t}.
\end{equation}

The average current to the lowest order in tunneling
\begin{equation}
 \mathrm{I}(t)=\sum_{\ell,\ell'=\mathcal{L},\mathcal{R}}\langle I_{\ell\ell'}(t)\rangle,
\label{general_current_formula}
\end{equation}
is the sum of four terms
\begin{equation}
\label{I_ll}
 I_{\ell\ell'}(t)=-\int_{-\infty}^t dt'I_{\ell\ell'}(t,t'),
 \end{equation}
with
\begin{equation}
I_{\ell\ell'}(t,t')=
\left[\widetilde{A}^{\dagger}_{\ell'}(t'),\widetilde{A}_\ell(t)\right]+
\left[\widetilde{A}^{\dagger}_{\ell'}(t),\widetilde{A}_\ell(t')\right].
\label{small_current_formula}
\end{equation}
The currents $\langle I_{\mathcal{L}\mathcal{L}}(t)\rangle$ and $\langle I_{\mathcal{R}\mathcal{R}}(t)\rangle$
are the direct terms at the left and right
tunnel junctions, respectively, while
$\langle I_{\mathcal{L}\mathcal{R}}(t)\rangle$, $\langle I_{\mathcal{R}\mathcal{L}}(t)\rangle$ are the interference terms.
And the brackets $\langle\,\cdots\rangle$ here mean averaging with respect to the state
\begin{equation}
\label{Initial_state}
|\Upsilon\rangle=\frac{|\tau_d|}{\sqrt{a}}\int_{-\infty}^0\!\!\! d\tau e^{i(\varepsilon_0-i\Gamma)\tau}
e^{{-\frac{i}{\sqrt{2}}}\sum_{\alpha}\xi_{\alpha u}(x_0+v_\alpha\tau)}|\Omega\rangle_u|\Omega\rangle_d.
\end{equation}
The time-dependent tunneling operator then reads
\begin{equation}
\widetilde{A}_\ell(t)=\tau_\ell\widetilde{\psi}_{1u}^\dag(x_\ell,t)\widetilde{\psi}_{1d}(x_\ell,t).
\end{equation}
The details of the calculation of the current are given in Appendices \ref{Sec:B} and \ref{Sec:C}.

\subsection{Direct current}
\label{Sec:IV}
From the expression for the direct current at the right tunnel junction
\begin{equation}
\label{direct_current}
\langle I_{\mathcal{R}\mathcal{R}}(t)\rangle=\frac{\Gamma |\tau_{\mathcal{R}}|^2}{4\pi^2uv}\sum_\alpha\left|\int_{-\infty}^0 d\tau \frac{e^{i(\varepsilon_0-i\Gamma)\tau}}{\tau+t+\frac{x_0-L}{v_\alpha}-i\gamma}\right|^2
\end{equation}
one can see that the charged and the dipole modes contribute independently to the direct current, i.e., tunneling of one mode does not affect
the tunneling of the other mode.
In the limit $\Gamma(t-|x_0-L|/v)\gg1$, Eq.\ (\ref{direct_current}) simplifies to
\begin{equation}
\label{direct_current_asympt}
\langle I_{\mathcal{R}\mathcal{R}}(t)\rangle=\frac{\Gamma|\tau_{\mathcal{R}}|^2}{uv}
\sum_\alpha\theta\left(t-\frac{|x_0-L|}{v_\alpha}\right)e^{-2\Gamma\left(t-\frac{|x_0-L|}{v_\alpha}\right)}.
\end{equation}

Integration of the current (\ref{direct_current}) over time $-\infty<t<\infty$ gives the total charge transmitted into the channel $1d$
through the right contact
\begin{equation}
\label{Q_RR}
Q_{\mathcal{R}\mathcal{R}}=\frac{|\tau_{\mathcal{R}}|^2}{2uv}\left[1+\frac2\pi\arctan\left(\frac{\varepsilon_0}{\Gamma}\right)\right].
\end{equation}
Interestingly, $Q_{\mathcal{R}\mathcal{R}}$ does not depend on the distance between the QD and the tunneling contact.
 Keeping in mind the approximation used in the derivation of the initial state for $\nu=2$, we are obligated to replace the second term in the brackets in Eq.\ (\ref{Q_RR}) by its value in the limit $\varepsilon_0/\Gamma\gg1$, therefore obtaining
\begin{equation}
\label{Q_RR_asymptotic}
Q_{\mathcal{R}\mathcal{R}}=\frac{|\tau_{\mathcal{R}}|^2}{uv}.
\end{equation}
Equivalently, the same result follows from the asymptotic expression (\ref{direct_current_asympt}) for the current, which indicates that we can safely use this asymptotic for the calculation of the transmitted charges with the required accuracy.

Finally, we recall that the expressions for the current $\langle I_{\mathcal{L}\mathcal{L}}(t)\rangle$ and the
associated charge $Q_{\mathcal{L}\mathcal{L}}$
can be obtained from the results (\ref{direct_current}-\ref{Q_RR_asymptotic}) by setting $L=0$ and replacing $\tau_{\mathcal{R}}$ with $\tau_{\mathcal{L}}$.
The total direct contribution to the transmitted charge in the limit $\varepsilon_0/\Gamma\gg1$ then reads
\begin{equation}
Q_{dir}=\int dt\left[\langle I_{\mathcal{L}\mathcal{L}}(t)\rangle+\langle I_{\mathcal{R}\mathcal{R}}(t)\rangle\right]=\frac{|\tau_{\mathcal{L}}|^2+|\tau_{\mathcal{R}}|^2}{uv}.
\end{equation}
We would like to mention that the free-fermionic case is recovered by setting $u=v=v_F$ in all the results presented above.

\subsection{Interference current}
\label{Sec:V}
Next, starting with the general expression for the interference contribution to the current
\begin{equation}
\mathrm{I}_{int}(t)=2\,\textrm{Re}[\langle I_{\mathcal{L}\mathcal{R}}(t)\rangle],
\end{equation}
and calculating it on the state (\ref{Initial_state}), we obtain
\begin{align}
\label{int_I_LR}
\langle I_{\mathcal{L}\mathcal{R}}(t)\rangle=-\frac{\Gamma\tau_\mathcal{L}\tau_\mathcal{R}^*}{2\pi^2uv\eta}
&\iint_{-\infty}^0 d\tau d\tau'\frac{e^{i\varepsilon_0(\tau-\tau')}e^{\Gamma(\tau+\tau')}}{\tau-\tau'-i\gamma}\nonumber \\ &\times[F(u,v)-F(v,u)],
\end{align}
where we have introduced the function
\begin{align}
&F(u,v)
= \nonumber\\ = &
\frac{\sqrt{-i(\tau^{\prime}+t+\frac{x_0}{v}-\frac{L}{v})+\gamma}\sqrt{i(\tau+t+\frac{x_0}{v}-\frac{L}{u})+\gamma}}
{\sqrt{-i(\tau^{\prime}+t+\frac{x_0}{v}-\frac{L}{u})+\gamma}\sqrt{i(\tau+t+\frac{x_0}{v}-\frac{L}{v})+\gamma}},
\end{align}
and the notation $\eta=\frac{L}{v}-\frac{L}{u}>0$.
Square roots in this function originate from the electron correlation functions derived with the help of the bosonization technique.

We note that neither the direct part nor the interference part of the transmitted charge depend on $x_0$.
For the interference part, this fact becomes obvious if one shifts the variable of integration $t\rightarrow t-x_0/v$.
Using Eq.~(\ref{int_I_LR}) one can present the formula for the charge as a sum of two terms:
the first term coincides with the transmitted charge in the case $x_0=0$, when the second term is zero.
Therefore, for simplicity we can set $x_0=0$:
\begin{equation}
\label{Interference_current}
\langle I_{\mathcal{L}\mathcal{R}}(t)\rangle=\frac{\Gamma\tau_\mathcal{L}\tau_\mathcal{R}^*}{2\pi^2uv}
\left|\int_{-\infty}^0\!\!\! d\tau \prod_\alpha\frac{e^{i(\varepsilon_0-i\Gamma)\tau/2}}{\sqrt{i(\tau+t-\frac{L}{v_\alpha})+\gamma}}\right|^2.
\end{equation}
Note that taking the limit $u,v\rightarrow v_F$ reduces this expression again to the free-fermion case.

Next, we rewrite Eq.\ (\ref{Interference_current}) in the different form
\begin{equation}
\label{Interference_current_Bessel}
\langle I_{\mathcal{L}\mathcal{R}}(t)\rangle=\frac{\Gamma\tau_\mathcal{L}\tau_\mathcal{R}^*}{2\pi^2uv}
\left|\int_0^\infty d\omega \frac{e^{-i(t-\lambda)\omega}J_0(\eta\omega/2)}{\omega-\varepsilon_0+i\Gamma}\right|^2,
\end{equation}
where $\lambda=\frac{L}{2v}+\frac{L}{2u}$, and $J_0(x)$ is the zeroth order Bessel function.
Using the formula (\ref{Interference_current_Bessel}), we obtain the interference contribution for the transmitted  charge,
\begin{equation}
Q_{int}=\frac{2\Gamma\mathrm{Re}(\tau_\mathcal{L}\tau_\mathcal{R}^*)}{\pi uv}
\int_0^\infty d\omega \frac{J_0^2(\eta\omega/2)}{(\omega-\varepsilon_0)^2+\Gamma^2}.
\end{equation}

Then, we consider the asymptotic behaviour of the current and transmitted charge in the limit $\Gamma\eta\ll1$, and for $t>\lambda$ we obtain
\begin{align}
\mathrm{I}_{int}(t)=&\frac{4\Gamma\,\mathrm{Re}(\tau_\mathcal{L}\tau_\mathcal{R}^*)}{uv}
J^2_0\left(\frac{\varepsilon_0\eta}{2}\right)e^{-2\Gamma\left(t-\lambda\right)},
\end{align}
and
\begin{equation}
Q_{int}=\frac{2\mathrm{Re}(\tau_\mathcal{L}\tau_\mathcal{R}^*)}{uv}J^2_0\left(\frac{\varepsilon_0\eta}{2}\right).
\end{equation}

In the opposite limit $\Gamma\eta\gg1$, which corresponds to strong dephasing, we obtain
\begin{align}
\mathrm{I}_{int}(t)=&\frac{4\Gamma\mathrm{Re}(\tau_\mathcal{L}\tau_\mathcal{R}^*)}{\pi uv\eta\varepsilon_0}\times \nonumber \\ \times&
\left|\sum_\alpha\theta\left(t-\frac{L}{v_\alpha}\right)
e^{-i\left(\varepsilon_0-i\Gamma\right)\left(t-\frac{L}{v_\alpha}\right)+i\vartheta_\alpha}\right|^2,
\end{align}
where $\vartheta_1=\pi/4$ and $\vartheta_2=-\pi/4$.
Interestingly, this result can be interpreted as interference of the amplitudes associated with charged and dipole mode.
The asymptotic expression for the transmitted charge then reads
\begin{equation}
Q_{int}=\frac{4\mathrm{Re}(\tau_\mathcal{L}\tau_\mathcal{R}^*)}{\pi uv}
\frac{1+e^{-\Gamma\eta}\sin{\varepsilon_0\eta}}{\varepsilon_0\eta}.
\end{equation}

\subsection{Visibility}
\label{Sec:VI}
The transmitted charge $Q=Q_{dir}+|Q_{int}|\cos(\sigma_{AB})$ oscillates as a function of the AB phase $\sigma_{AB}=\mathrm{Arg}(\tau_\mathcal{L}\tau_\mathcal{R}^*)$ between the maximal $Q_{max}$ and the minimal $Q_{min}$ values.
The degree of phase coherence is described by the visibility of these oscillations:
\begin{equation}
V=\frac{Q_{max}-Q_{min}}{Q_{max}+Q_{min}}=\frac{|Q_{int}|}{Q_{dir}}.
\end{equation}
For the case of free-fermions this formula reproduces the well-known result
\begin{equation}
V_0=\frac{2|\tau_\mathcal{L}\tau_\mathcal{R}|}{|\tau_\mathcal{L}|^2+|\tau_\mathcal{R}|^2}.
\end{equation}
We note that the same result may be as well obtained from the general expressions for $Q_{dir}$ and $Q_{int}$ in the limit of $\varepsilon_0\eta\rightarrow0$,
which is very natural, since in this limit there is no sufficient energy is available in the initial state for dephasing to take place.

\begin{figure}[]
\includegraphics[width=8.5cm]{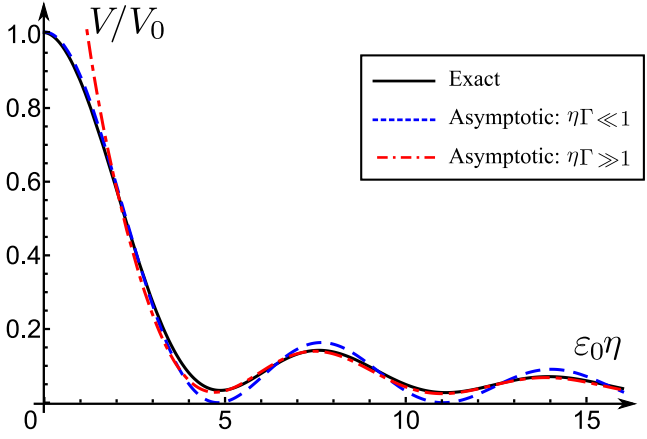}
\caption{(Color online) The normalized visibility $V/V_0$ as a function of
$\varepsilon_0\eta$ for the case $\varepsilon_0/\Gamma=20$.}
\label{Visibility}
\end{figure}

Using the results of Secs.\ \ref{Sec:IV} and \ref{Sec:V}, we find the analytical expressions for the visibility in the two limits:
\begin{equation}
V=J^2_0\left(\frac{\varepsilon_0\eta}{2}\right)V_0, \quad \Gamma\eta\ll1;
\label{visibility-one}
\end{equation}
\begin{equation}
V=\frac{2}{\pi}\frac{1+e^{-\Gamma\eta}\sin{\varepsilon_0\eta}}{\varepsilon_0\eta}V_0, \quad \Gamma\eta\gg1.
\label{visibility-two}
\end{equation}
We see that the visibility vanishes at $\varepsilon_0\eta\gg1$.
The exact and asymptotic results for the visibility are presented in Fig.\ \ref{Visibility}.

To conclude this section, we would like to emphasize the two most surprising aspects of our results:
({\it{i}}) In contrast to the experiment [\onlinecite{Tewari}], where the saturation of the visibility of the AB oscillations
has been observed for energies of a wave packet larger than certain threshold  energy, in our model we find a power-law decay of the visibility as a function of energy;
({\it{ii}}) Even more startling is the fact that the visibility, according to our calculations, does not depend on the distance from the injecting QD, despite the fact that the interaction gradually splits the wave packets into the dipole and the charged modes right after the injection. In order to clarify this puzzling phenomenon, in the following section we investigate the Wigner representation of the wave packet and find its asymptotic form at long distances.

\section{Wigner function}
\label{Sec:VII}
The Wigner function of the electron excitations in the channel $\alpha$ of the edge at $\nu=2$ is defined as
\begin{equation}
W_\alpha(x,\varepsilon,t)=\!\!
\int dz \frac{e^{-i\varepsilon z}}{2\pi}
\langle\psi_\alpha^\dag\Big(x,t+\frac{z}{2}\Big)\psi_\alpha\Big(x+\delta,t-\frac{z}{2}\Big)\rangle,
\end{equation}
where the point-splitting parameter $\delta$ is to be set to zero after the integration over $z$.
This procedure allows one to subtract correctly the contribution
of the Fermi sea $W_\alpha^{FS}(x,\varepsilon)$ to the Wigner function.
The difference
\begin{equation}
\Delta W_\alpha(x,\varepsilon,t)=W_\alpha(x,\varepsilon,t)-W_\alpha^{FS}(x,\varepsilon,t)
\end{equation}
originates from the injected electron and is an experimentally measurable quantity.\cite{Jullien}
It is normalized to the number $N_\alpha$ of electrons above the Fermi sea in the channel $\alpha$,
\begin{equation}
\iint dx d\varepsilon \Delta W_\alpha(x,\varepsilon,t)=N_\alpha.
\end{equation}
In our case, $N_1=1$ and $N_2=0$, because only one electron is injected to the first channel,
and there is no tunneling between the first and the second channel.
This Wigner function can be used to evaluate the induced excess particle density
\begin{equation}
\Delta\rho_\alpha(x,t)=\int d\varepsilon \Delta W_\alpha(x,\varepsilon,t),
\end{equation}
and the energy distribution of the injected electron
\begin{equation}
\Delta f_\alpha(\varepsilon,t)=\int dx \Delta W_\alpha(x,\varepsilon,t).
\end{equation}

Subtracting the contribution of the Fermi sea from the initial state, we present $\Delta W_\alpha(x,\varepsilon)$ in the following form:
\begin{align}
\label{WF_1}
\Delta W_1(x,\varepsilon,t)=&-\frac{\Gamma}{4\pi^3}\iint_{-\infty}^0 d\tau d\tau'
\frac{e^{i\varepsilon_0(\tau-\tau')}e^{\Gamma(\tau+\tau')}
}{\tau-\tau'-i\gamma}\nonumber\\ \times&\int dz
\frac{e^{-i\varepsilon z}[\chi(u,\delta)\chi(v,\delta)-1]}{\sqrt{uz+\delta-i\gamma}\sqrt{vz+\delta-i\gamma}},
\end{align}
\begin{align}
\label{WF_2}
\Delta W_2(x,\varepsilon,t)=&-\frac{\Gamma}{4\pi^3}\iint_{-\infty}^0 d\tau d\tau'
\frac{e^{i\varepsilon_0(\tau-\tau')}e^{\Gamma(\tau+\tau')}
}{\tau-\tau'-i\gamma}\nonumber\\ \times&\int dz
\frac{e^{-i\varepsilon z}[\chi(u,\delta)\chi^{-1}(v,\delta)-1]}{\sqrt{uz+\delta-i\gamma}\sqrt{vz+\delta-i\gamma}},
\end{align}
where
\begin{align}
\label{chi}
\chi(v_\alpha,\delta)=&\sqrt{\frac{x-x_0-v_\alpha t-v_\alpha z/2-v_\alpha\tau+i\gamma}{\delta+x-x_0-v_\alpha t+v_\alpha z/2-v_\alpha\tau+i\gamma}}\nonumber\\ \times&\sqrt{\frac{\delta+x-x_0-v_\alpha t+v_\alpha z/2-v_\alpha\tau'-i\gamma}{x-x_0-v_\alpha t-v_\alpha z/2-v_\alpha\tau'-i\gamma}}.
\end{align}

First, we calculate the particle density by integrating $\Delta W_\alpha(x,\varepsilon,t)$ over $\varepsilon$,
then over $z$, and taking the limit $\delta\rightarrow0$
\begin{equation}
\Delta\rho_\alpha(x,t)=h_1(x,t)+(-1)^{\alpha-1}h_2(x,t),
\end{equation}
where
\begin{equation}\label{h_alpha}
h_\alpha(x,t)=\frac{\Gamma}{4\pi^2v_\alpha}\left|\int_{-\infty}^0\!\!\!d\tau
\frac{e^{i(\varepsilon_0-i\Gamma)\tau}}{\tau+t-\frac{x-x_0}{v_\alpha}-i\gamma}\right|^2 .
\end{equation}
Again, the same result can also be obtained by averaging the time-dependent density operators expressed directly in terms of the  bosonic fields
$\Delta\rho_\alpha(x,t)=\langle\partial_x\phi_\alpha(x,t)\rangle/2\pi$.
The asymptotic form for $h_\alpha(x,t)$ in the limit $\varepsilon_0/\Gamma\gg1$ reads
\begin{equation}
h_\alpha(x,t)=\frac{\Gamma}{v_\alpha}\theta\Big(t-\frac{x-x_0}{v_\alpha}\Big)
e^{-2\Gamma\big(t-\frac{x-x_0}{v_\alpha}\big)}.
\end{equation}
And the normalization $N_1=1$ and $N_2=0$ can be easily verified.

As one can see, the charged and the dipole excitations contribute independently to the particle density of each channel.
Moreover, at $t\rightarrow\infty$ they are spatially separated.
Accordingly, $\Delta W_\alpha(x,\varepsilon,t)$ can be presented as a sum of the
contributions of charged and dipole modes in the long-time limit.
We apply this observation to Eq.\ (\ref{WF_1}).
The functions $\chi(v,\delta)$ and $\chi(u,\delta)$ represent the wave packets of the dipole and the charged mode,
respectively.
Therefore, the product $\chi(u,\delta)\chi(v,\delta)-1$ may be split as a sum of $\chi(u,\delta)-1$ and $\chi(v,\delta)-1$
in the limit $t\rightarrow\infty$, where the modes are spatially separated. In this case one obtains
\begin{equation}
\Delta W_1(x,\varepsilon,t)=\sum_\alpha w_\alpha(x,\varepsilon),
\end{equation}
with
\begin{align}
\label{w_alpha}
w_\alpha(x,\varepsilon)=&-\frac{\Gamma}{4\pi^3}\iint_{-\infty}^0 d\tau d\tau'
\frac{e^{i\varepsilon_0(\tau-\tau')}e^{\Gamma(\tau+\tau')}
}{\tau-\tau'-i\gamma}\nonumber\\ \times&\int dz
\frac{e^{-i\varepsilon z}[\chi(v_\alpha,\delta)-1]}{\sqrt{uz+\delta-i\gamma}\sqrt{vz+\delta-i\gamma}}.
\end{align}
As a result of this splitting, the energy distribution function also consists of
the charged and the dipole modes' contributions
 \begin{align}
 \Delta f_1(\varepsilon,t)=\sum_\alpha f_\alpha(\varepsilon).
\end{align}
It is convenient to present these distributions as a sum of an odd and an even functions of energy
\begin{equation}
\label{e_o_separation}
w_\alpha(x,\varepsilon)=w_\alpha^e(x,\varepsilon)+w_\alpha^o(x,\varepsilon),
\end{equation}
\begin{equation}
f_\alpha(\varepsilon)=f_\alpha^e(\varepsilon)+f_\alpha^o(\varepsilon).
\end{equation}

Before proceeding with the calculations, we would like to comment on the role of
the point-splitting parameter $\delta$ for the calculation of the Wigner function.
Let us consider (\ref{w_alpha}) as a function of $\varepsilon$.
The low-energy dependence of it comes from large $z$, where there is no significant
difference between the integrand with zero $\delta$ and small nonzero values of $\delta$. Therefore, Eq.\ (\ref{w_alpha}) with $\delta=0$ describes the low-energy behavior of the Wigner function. Further details of the point-splitting procedure are discussed in Appendix \ref{Sec:D}.
\begin{figure}[tl]
\includegraphics[width=8.5cm]{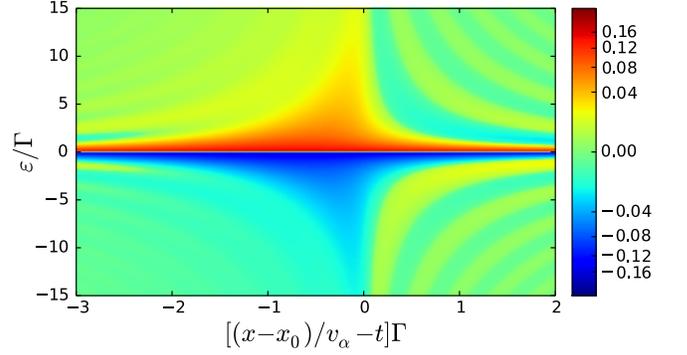}
\caption{(Color online) The density plot of the normalized odd part of the Wigner function $\sqrt{uv}w_\alpha^o(x,\varepsilon)$
versus dimensionless coordinates $[(x-x_0)/v_\alpha-t]\Gamma$ and $\varepsilon/\Gamma$ is shown. $w_\alpha^o(x,\varepsilon)$ takes
positive and negative values in upper and lower half-plane, respectively.}
\label{WignerOdd}
\end{figure}

Let's consider the odd part of the Wigner function at low energies in the limit $\varepsilon_0/\Gamma\gg1$.
The main contribution to this term comes from the region of the width $\varepsilon_0^{-1}$, close to the line $\tau=\tau'$.
Taking this fact into account, one can evaluate the integrals in (\ref{w_alpha}) asymptotically
\begin{equation}
\label{w_alpha_odd}
w_\alpha^o(x,\varepsilon)=\frac{2\Gamma sgn\varepsilon}{\pi^2\sqrt{uv}}\int_{-\infty}^0\!\!\! d\tau e^{2\Gamma\tau} \left[\frac{\pi}{2}-Si(2|\tau-\tau_\alpha||\varepsilon|)\right],
\end{equation}
where $\tau_\alpha=(x-x_0)/v_\alpha-t$ and $Si(x)$ is the Sine integral.
This result is only applicable in the region of energies $|\varepsilon|\leq \varepsilon_0$
due to the limitations of the computational method, used here.
We also note that the obtained formula is only valid in the case of complete spatial separation of the charged and the dipole modes.
This means that the expression (\ref{w_alpha_odd}) does not apply at energies $\varepsilon<(u+v)/2\pi(u-v)t$.

The density plot for the odd part of Wigner function, that describes electron-hole pairs, excited by the electron injected with the
energy $\varepsilon_0$, is presented on Fig.~\ref{WignerOdd}.
\begin{figure}[t]
\includegraphics[width=8.5cm]{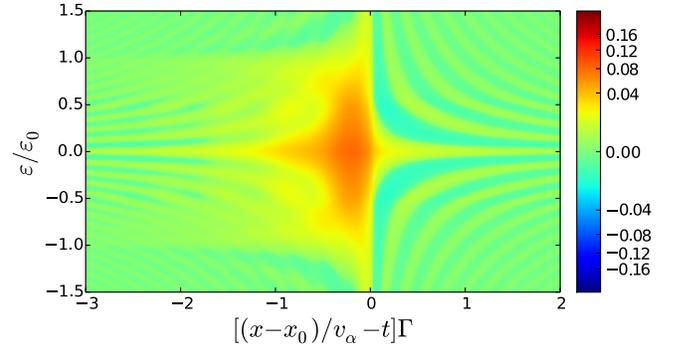}
\caption{(Color online) The color plot of the normalized even part of the Wigner function $\sqrt{uv}w_\alpha^e(x,\varepsilon)$ versus dimensionless coordinate $[(x-x_0)/v_\alpha-t]\Gamma$ and the normalized energy $\varepsilon/\varepsilon_0$ is shown. Here $\varepsilon_0/\Gamma=10$.}
\label{WignerEven}
\end{figure}
The corresponding odd component of the energy distribution function is
\begin{align}
\label{f alpha o}
f_\alpha^o(\varepsilon)=\frac{v_\alpha}{\pi^2\sqrt{uv}\varepsilon}.
\end{align}
Indeed, the contribution of (\ref{f alpha o}) to the average number of particles is zero, while its contribution to the total
energy is close to $\varepsilon_0$.

The density plot for the symmetric low-energy part of the Wigner function
\begin{align}
w_\alpha^e(x,\varepsilon)=&\frac{\Gamma}{8\pi^3\sqrt{uv}}
\int dze^{-i\varepsilon z}\\ \nonumber \times&
\left|\int_0^\infty d \omega \frac{e^{i(t-\frac{x-x_0}{v_\alpha})\omega}}
{\omega-\varepsilon_0+i\Gamma}J_0\left(\frac{z\omega}{2}\right)\right|^2,
\end{align}
corresponding to the mode with velocity $v_\alpha$, in the limit of $t\rightarrow\infty$
is presented in Fig.\ \ref{WignerEven}.
The even part of the energy distribution function can be presented in the integral form
\begin{align}
f_\alpha^e(\varepsilon)=&\frac{\Gamma v_\alpha}{4\pi^2\sqrt{uv}}
\int_0^\infty d\omega \int  dz
\frac{e^{-i\varepsilon z}J_0^2\left(\frac{z \omega}{2}\right)}{(\omega-\varepsilon_0)^2+\Gamma^2},
\end{align}
which can be evaluated in the limit $\Gamma/\varepsilon_0\to 0$, as
\begin{equation}
\label{Asymptotic}
f_\alpha^e(\varepsilon)=\frac{v_\alpha}{\pi^2\sqrt{uv}|\varepsilon|}K\left(1-\frac{\varepsilon_0^2}{\varepsilon^2}\right)
\theta\left(1-\frac{\varepsilon^2}{\varepsilon_0^2}\right),
\end{equation}
where $K(x)$ is the complete elliptic integral of the first kind, and the $\theta$-function truncates it at $|\varepsilon|=\varepsilon_0$.
This function has the asymptotics
\begin{equation}
f_\alpha^e(\varepsilon)=\frac{v_\alpha}{\pi^2\sqrt{uv}\varepsilon_0}\ln\left(\frac{4\varepsilon_0}{|\varepsilon|}\right)
,\quad \varepsilon/\varepsilon_0\ll1;
\end{equation}
\begin{equation}
f_\alpha^e(\varepsilon)=\frac{v_\alpha}{2\pi\sqrt{uv}\varepsilon_0}
\theta\left(1-\frac{\varepsilon^2}{\varepsilon_0^2}\right),\quad \varepsilon\rightarrow\varepsilon_0.
\end{equation}

\begin{figure}[]
\includegraphics[width=8.5cm]{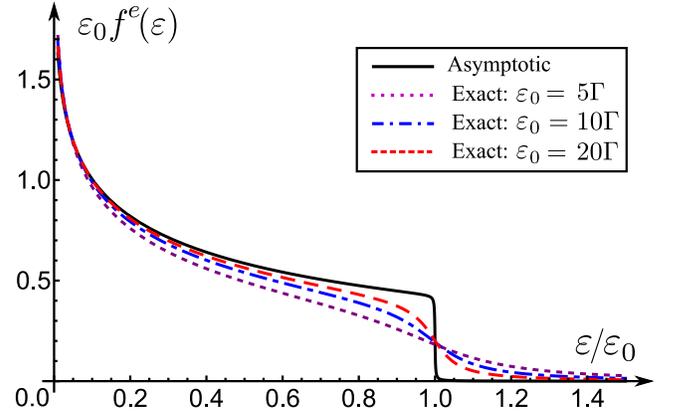}
\caption{(Color online) The even part of the normalized energy distribution
function $\varepsilon_0f^e(\varepsilon)$ as a function of normalized
$\varepsilon/\varepsilon_0$ for $\varepsilon>0$ and $u=5v$. The black curve describes
analytical result (\ref{Asymptotic}), valid for $\varepsilon_0/\Gamma\gg1$. The purple
dotted, blue dot-dashed, and red dashed curves represent exact results for the
case $\varepsilon_0/\Gamma=5,10,20$, respectively.}
\label{DistributionSymm}
\end{figure}

The total even part of the energy distribution function $f^e(\varepsilon)=f_1^e(\varepsilon)+f_2^e(\varepsilon)$ for different values of the parameters is presented in Fig.\ \ref{DistributionSymm}. By integrating this function over $\varepsilon$, one finds that the total number of particles in the wave-packet is not equal to one.  This is because the remaining contribution comes from the correction to the integral (\ref{w_alpha}) at high energies, and does not affect the distribution at moderate energies. Indeed, according to the point-splitting procedure, this correction
 $\varpi_\alpha(x,\varepsilon)=w_\alpha(x,\varepsilon)-w_\alpha(x,\varepsilon)|_{\delta=0}$
 becomes important at small $z\sim\delta$ in (\ref{w_alpha}). This means that one can expand the numerator of the integrand in (\ref{w_alpha})
as a series in powers of $z$ and $\delta$. Then the correction takes the form
\begin{align}
\label{hecorrection}
\varpi_\alpha(x,\varepsilon)=&\frac{\Gamma}{8\pi^3v_\alpha}\left|\int_{-\infty}^0 d\tau
\frac{e^{i(\varepsilon_0-i\Gamma)\tau}
}{t+\tau-\frac{x-x_0}{v_\alpha}-i\gamma}\right|^2\nonumber \\&\times \int dz
e^{-i\varepsilon z}[\zeta_\alpha(\delta)-\zeta_\alpha(0)],
\end{align}
where
\begin{equation}
\zeta_\alpha(\delta)=\frac{\delta+v_\alpha z}{\sqrt{uz+\delta-i\gamma}\sqrt{vz+\delta-i\gamma}}.
\end{equation}

 Note, that this correction is a product of the spatially- and energy-dependent functions. The spatial term reproduces the
 particle density profile of the injected electron (\ref{h_alpha}), while the energy-dependent part becomes relevant only in the region   $\varepsilon<-\delta^{-1}$.
 On the other hand, the energy of electrons in a physical system is restricted by the bottom of the electronic band.
 Therefore, the injection of an electron into the channel must induce a deformation  of the bottom of
 the band, which determines the high-energy correction.
\begin{figure}[]
\includegraphics[width=8.5cm]{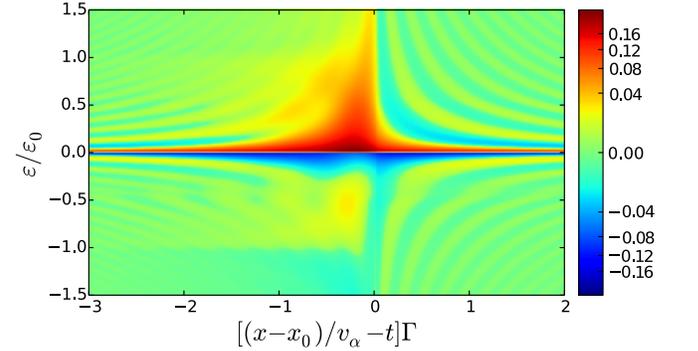}
\caption{(Color online) The color plot of the normalized Wigner function $\sqrt{uv}w_\alpha(x,\varepsilon)$ as a function of
 coordinate $[(x-x_0)/v_\alpha-t]\Gamma$ and normalized energy $\varepsilon/\varepsilon_0$. Here $\varepsilon_0/\Gamma=10$.}
\label{total}
\end{figure}

To summarize this section, we recap the most important stages of the evolution of the state of the injected electron.
For the short times $\Gamma t\ll1$, right after the moment of the injection of an electron into the channel, the two modes start to form.
The energy distribution function of the corresponding state has a Lorentzian profile with the energy $\varepsilon_0$ and the width $\Gamma$.
After the time $\Gamma t\sim1$ the modes are spatially separated. The Coulomb interaction smears the initial energy distribution function over the
region $|\varepsilon|\leq\varepsilon_0$.
In the long-time limit $\Gamma t\gg1$ the modes behave independently. The Wigner function of each wave-packet does not change with time,
and is characterized by the parameters of the injected electron, namely by $\varepsilon_0$ and $\Gamma$, see Fig.\ \ref{total}.
Thus, as a result of the integrability of the system, the final steady state retains the memory of the initial state.
Interactions, no matter how strong they are, do not lead to a complete relaxation of the initial state.
Therefore, the visibility vanishes with the increasing energy of the injected electron.

\section{Conclusion}
\label{Sec:VIII}

Motivated by the puzzling results of the recent experiment from the group of Patrice Roche [\onlinecite{Tewari}], we have investigated the evolution of a single-electron wave packet in chiral one-dimensional electron channels on the edge of a QH system at filling factor $\nu=2$. The experimentalists have used  resonant tunnelling of electrons via a QD, tunnel-coupled to the electron channel, in order to create such states, and a MZ interferometer embedded downstream of the quantum dot to detect the degree of phase coherence remaining after the strong Coulomb interaction between channels has considerably affected the state. The question that arises in the context of this physical situation is whether the strong interactions are able to relax the initial state, created at the energy $\varepsilon = \varepsilon_0$, down to the Fermi level $\varepsilon=0$ of one-dimensional electrons of the edge channels, as they typically do for electrons in Fermi-liquid systems, or whether the free-bosonic character of the collective charge excitations, resulting in the trivial integrability of the system, does not allow this to happen.

In order to investigate this problem, we have applied the model of interacting edge states at filling factor $\nu=2$, proposed earlier in Ref.\ \onlinecite{Levkivskyi_1}. This model has been successfully used for the theoretical explanation of a number of recent mesoscopic experiments with QH edge states.
\cite{Neder_1,Litvin_1, Preden, Neder_2, Litvin_2,Bieri, Huynh} According to this model, tunnelling of an electron to chiral edge channels is not affected by the interaction and can be described exactly using free-fermionic approach. At the same time, the strong interactions that affect the state after the injection, are non-perturbative, although they can be analysed exactly using the free-bosonic approach. Using this technique, we have  evaluated the visibility of the AB oscillations and have found that it decays with the energy $\varepsilon_0$ of the initial state as a power-law function [see Eqs.\ (\ref{visibility-one}) and (\ref{visibility-two})], contrary to the observation in Ref.\ [\onlinecite{Tewari}] where it saturates at large energies. This indicates that despite strong interactions, that considerably deform the initial state before it reaches the interferometer, it still remembers the initial conditions and in particular the energy $\varepsilon_0$ of the injected state.

In order to further clarify the mechanisms of the relaxation of the initial state towards the quasi-stationary state at long distances, where the interaction splits the electron wave packet into the dipole and the charged modes, we follow the Refs.\ [\onlinecite{Ferraro_1,Ferraro_2}] and investigate the electron distribution function and the Wigner function of the outgoing state. We find that the effects of the propagation and of the strong interactions is twofold: they distribute an electron, initially created with energy $\varepsilon_0$ above the Fermi sea, almost homogeneously inside the interval of energies from $-\varepsilon_0$ to $\varepsilon_0$, and create a bunch of electron-hole excitations around the Fermi level distributed as $1/\varepsilon$ with the almost sharp cutoff at $|\varepsilon|=\varepsilon_0$ [see Eqs.\ (\ref{f alpha o}) and (\ref{Asymptotic})]. This result helps us to understand the behaviour of the visibility as a function of the initial energy. Indeed, taking into account the fact that the linear dispersion law, assumed for the electrons in our model, conserves the width of the wave packets, the total injected energy $\varepsilon_0$ is locked inside the finite space interval and this heats the state up to the (effective) temperatures of the order of $\varepsilon_0$. As a result, the coherence of the state is suppressed inside the interferometer, depending on the initial energy $\varepsilon_0$.

An additional interesting observation is that the excess distribution function $f(\varepsilon)-\theta(-\varepsilon)$ that describes the contribution of the injected electron, does not carry the complete charge of one electron (in other words, its integral over energy is not equal to 1).
However, the point-splitting procedure restores the total charge of the injected electron. This indicates that the additional contribution to the total electron charge arises from the high energies and cannot be directly accounted by the bosonization procedure, which relies on the linearization of the spectrum of electrons. Taking into account the incompressibility of the Fermi sea, the remaining charge can only be attributed to the interaction-induced local deformation of the bottom of the conduction band, which propagates with the wave packets. We support this conclusion by considering an additional example of this effect in Appendix \ref{Sec:D}.

Finally, we would like to mention that a similar problem of relaxation of an initial state towards a non-equilibrium quasi-stationary state at the edge of a QH system has been considered in Refs.\ [\onlinecite{Levkivskyi_2,Levkivskyi_3}]. However, as the initial state, the authors of these works have considered a double-step electron distribution created by voltage biasing a QPC connecting the two channels. They have found that the excess distribution also scales as $1/\varepsilon$ and is cutoff at the energies of the order of the voltage bias $eV$. In addition, for the case of low QPC transparencies ${\cal T}\ll 1$, the distribution acquires the low-energy cutoff at $\varepsilon\sim eV{\cal T}$. Taking into account the results of the present paper, the conclusion is rather obvious: The electron distribution in this case arises from the individual contributions of the electrons injected trough the QPC in the energy interval from $0$ to $eV$, accompanied with the effect of partition noise, that cut off the distribution at low energies.

It remains to understand the puzzling results of the experiment [\onlinecite{Tewari}]. The saturation of the visibility at finite values points at an additional physical effect arising at certain energies, which is not accounted for our model, and which cannot be explained solely by strong interactions. 
We can only speculate that this phenomenon requires accounting for the non-linearities in the spectrum of electrons, and perhaps considering the strong effects of the disorder which is always present at a QH edge.

\acknowledgments
We thank  Iu.\ Chernii, D.\ Ferraro, I.\ Levkivskyi for fruitful discussions.
This work was supported by the Swiss National Science
Foundation.

\appendix

\section{Derivation of the initial state}
\label{Sec:A}
As we have argued in Sec. \ref{Sec:II}, an electron tunneling from the QD creates the following free-fermion state
\begin{equation}
|\Psi\rangle_{in}^F=\frac{|\tau_d|}{v_F}\int_{x_0-v_FT}^{x_0} dx f(x)\psi_1^\dag(x)|\Omega\rangle
\end{equation}
as the initial state. Here, the state
$|\Omega\rangle=|0\rangle_1|0\rangle_2$ is the ground state of the Hamiltonians $H_1$ (\ref{H_1}), and $H_2$ (\ref{H_2})
and
\begin{equation}
\label{wave-packet}
f(x)=e^{i(\varepsilon_0-i\Gamma)(x-x_0)/v_F}
\end{equation}
 is the free-fermionic single particle wave function.
Evolution of this state during time $T$, generated by the Hamiltonian $H_1$, results in the state
\begin{align}
\label{Psi_out}
|\Psi\rangle_{out}=&e^{-iH_1T}|\Psi\rangle_{in}^F=\frac{|\tau_d|}{v_F}\int_{x_0-v_FT}^{x_0} dx f(x)\times \nonumber \\ \times &e^{-iH_1T}\psi_1^\dag(x)e^{iH_1T}|\Omega\rangle.
\end{align}
Taking into account the bosonized representation of the electronic operator
\begin{equation}
\psi_1^\dag(x)=\frac{1}{\sqrt{a}}e^{-\frac{i}{\sqrt{2}}[\xi_1(x)+\xi_2(x)]},
\end{equation}
one needs to find the functions
\begin{equation}
\xi_\alpha(x,t)=e^{-iH_1t}\xi_\alpha(x)e^{iH_1t},
\end{equation}
that are the solutions of the equation of motion reads
\begin{align}
\partial_t\xi_\alpha(x,t)=&-ie^{-iH_1t}[H_1,\xi_\alpha(x)]e^{iH_1t}=\nonumber \\ =&
\left\{
\begin{array}{c}
v_F\partial_x\xi_\alpha(x,t), \quad x<x_0; \\
v_\alpha\partial_x\xi_\alpha(x,t), \quad   x>x_0. \\
\end{array}\right.
\end{align}
and may be presented in the form
\begin{equation}
\xi_\alpha(x,t)=\xi_\alpha(x+v_Ft)
\end{equation}
for $ x<x_0-v_F t$. Then
\begin{equation}
\xi_\alpha(x,t)=\xi_\alpha(x_0+v_\alpha[x+v_Ft-x_0]/v_F)
\end{equation}
for the interval $[x_0-v_F t,x_0]$, and finally
\begin{equation}
\xi_\alpha(x,t)=\xi_\alpha(x+v_\alpha t)
\end{equation}
for $x>x_0$.
We substitute this result to Eq.\ (\ref{Psi_out}) and find the evolution of the obtained state
backward in time with the Hamiltonian $H_2$,
\begin{equation}
|\Psi\rangle_{in}^B=e^{iH_2T}|\Psi\rangle_{out}.
\end{equation}
After substituting $x=x_0+v_F\tau$, the state takes the form
\begin{align}
|\Psi\rangle_{in}^B=\frac{|\tau_d|}{\sqrt{a}}\int_{-T}^0
d\tau f(x_0+v_F\tau)e^{-\frac{i}{\sqrt{2}}\sum_\alpha\xi_\alpha(x_0+v_\alpha\tau)}|\Omega\rangle.
\end{align}
Using the explicit form of the wave packet (\ref{wave-packet}), we thus arrive to the expression (\ref{Psi_in^B}) for the initial state.

\section{Derivation of the direct contribution to the current}
\label{Sec:B}
To obtain this expression, we use the local anticommutation relation for the time-dependent fermion fields.
For Eq.\ (\ref{small_current_formula}), we need to calculate the commutator
\begin{equation}
\mathcal{G}(t',t)=\left[\widetilde{A}^\dagger_\mathcal{R}(t'),\widetilde{A}_\mathcal{R}(t)\right],
\end{equation}
which we present in terms of fermion fields
\begin{equation}
\mathcal{G}(t',t)=|\tau_{\mathcal{R}}|^2\left[\widetilde{\psi}_{1d}^\dag(L,t')\widetilde{\psi}_{1u}(L,t'),
\widetilde{\psi}_{1u}^\dag(L,t)\widetilde{\psi}_{1d}(L,t)\right],
\end{equation}
and arrive to the expression
\begin{align}
\mathcal{G}(t',t)&=|\tau_{\mathcal{R}}|^2\left[\widetilde{\psi}_{1d}^\dag(L,t^{\prime})\widetilde{\psi}_{1d}(L,t)\left\{\widetilde{\psi}_{1u}(L,t'),
\widetilde{\psi}_{1u}^\dag(L,t)\right\}\right.
 \nonumber\\ &-\left.
\widetilde{\psi}_{1u}^\dag(L,t)\widetilde{\psi}_{1u}(L,t')\left\{\widetilde{\psi}_{1d}(L,t),\widetilde{\psi}_{1d}^\dag(L,t')\right\}\right].
\end{align}

The latter can be simplified with the help of the local anticommutation relation for time-dependent fermionic operators:
\begin{equation}
\{\widetilde{\psi}_{1j}(x,t^{\prime}),\widetilde{\psi}_{1j}^\dag(x,t)\}=\frac{\delta(t^{\prime}-t)}{\sqrt{uv}},
\end{equation}
where
\begin{equation}
\widetilde{\psi}_{1j}(x,t)=\frac{1}{\sqrt{a}}e^{\frac{i}{\sqrt{2}}[\xi_{1j}(x-ut)+\xi_{2j}(x-vt)]}.
\end{equation}
Namely, one can write
\begin{equation}
\label{G}
\mathcal{G}(t',t)=|\tau_{\mathcal{R}}|^2\mathcal{K}(t',t)\frac{\delta(t-t')}{\sqrt{uv}},
\end{equation}
where
\begin{equation}
\label{K}
\mathcal{K}(t',t)=\widetilde{\psi}_{1d}^\dag(L,t^{\prime})\widetilde{\psi}_{1d}(L,t)-\widetilde{\psi}_{1u}^\dag(L,t)\widetilde{\psi}_{1u}(L,t^{\prime}).
\end{equation}
Using Eqs.\ (\ref{G}) and (\ref{K}), we arrive at the expression for the direct current
\begin{equation}
\label{I_RR}
\langle I_{\mathcal{R}\mathcal{R}}(t)\rangle=-\frac{|\tau_{R}|^{2}}{\sqrt{uv}}\int^{t}_{-\infty}dt'\delta(t-t')
\langle\mathcal{K}(t',t)+\mathcal{K}(t,t')\rangle.
\end{equation}
where the average is with respect to the initial state (\ref{Initial_state}).
The total correlation function on the right hand side splits into the product of independent correlators
\begin{align}
\label{d_correlator}
\langle\widetilde{\psi}_{1d}^\dag(L,t')&\widetilde{\psi}_{1d}(L,t)\rangle\!=\!
\frac{|\tau_d|^2}{a^2}\!\iint_{-\infty}^0\!\!\!\! d\tau d\tau' e^{i\varepsilon_0(\tau-\tau')}e^{\Gamma(\tau+\tau')}\nonumber \\ \times
\prod_\alpha&\langle0|e^{\frac{i}{\sqrt{2}}\xi_{\alpha u}(x_0+v_\alpha\tau')}
e^{-\frac{i}{\sqrt{2}}\xi_{\alpha u}(x_0+v_\alpha\tau)}|0\rangle_{\alpha u} \nonumber \\ \times
&\langle0|e^{-\frac{i}{\sqrt{2}}\xi_{\alpha d}(L-v_\alpha t')} e^{\frac{i}{\sqrt{2}}\xi_{\alpha d}(L-v_\alpha t)}|0\rangle_{\alpha d},
\end{align}
and
\begin{align}
\langle\widetilde{\psi}_{1u}^\dag(L,t)&\widetilde{\psi}_{1u}(L,t')\rangle\!=\!
\frac{|\tau_d|^2}{a^2}\!\iint_{-\infty}^0 \!\!\!\! d\tau d\tau' e^{i\varepsilon_0(\tau-\tau')}e^{\Gamma(\tau+\tau')} \nonumber \\ \times
\prod_\alpha&\langle0|e^{\frac{i}{\sqrt{2}}\xi_{\alpha u}(x_0+v_\alpha\tau')}
e^{-\frac{i}{\sqrt{2}}\xi_{\alpha u}(L-v_\alpha t)} \nonumber \\ &\times
e^{\frac{i}{\sqrt{2}}\xi_{\alpha u}(L-v_\alpha t')}
e^{-\frac{i}{\sqrt{2}}\xi_{\alpha u}(x_0+v_\alpha\tau)}|0\rangle_{\alpha u}.
\end{align}
Averaging them over the fluctuations of the bosonic fields, we obtain
\begin{equation}
\label{2_correlator}
\langle0|e^{\pm\frac{i}{\sqrt{2}}\xi_{\alpha j}(x)}
e^{\mp\frac{i}{\sqrt{2}}\xi_{\alpha j}(y)}|0\rangle_{\alpha j}
=\frac{\sqrt{\gamma}}{\sqrt{i(y-x)+\gamma}},
\end{equation}
and
\begin{align}
\label{4_correlator}
&\langle0|e^{\frac{i}{\sqrt{2}}\xi_{\alpha j}(x)}
e^{-\frac{i}{\sqrt{2}}\xi_{\alpha j}(y)}e^{\frac{i}{\sqrt{2}}\xi_{\alpha j}(z)}
e^{-\frac{i}{\sqrt{2}}\xi_{\alpha j}(t)}|0\rangle_{\alpha j}=\nonumber \\ &
\frac{\gamma \sqrt{i(z-x)+\gamma}\sqrt{i(t-y)+\gamma}}{\sqrt{i(y-x)+\gamma}\sqrt{i(z-y)+\gamma}\sqrt{i(t-z)+\gamma}\sqrt{i(t-x)+\gamma}},
\end{align}
where $\gamma$ is an infinitesimally small parameter, chosen so that the last relation can be obtained by comparing it
to the free-fermion correlators.
Substituting Eqs.\ (\ref{d_correlator}-\ref{4_correlator}) to Eq.\ (\ref{I_RR}), we obtain
\begin{equation}
\langle I_{\mathcal{R}\mathcal{R}}(t)\rangle=\frac{\Gamma|\tau_{\mathcal{R}}|^2}{4\pi^2uv}
\sum_\alpha\left|\int_{-\infty}^0 d\tau \frac{e^{i(\varepsilon_0-i\Gamma)\tau}}{\tau+t+\frac{x_0-L}{v_\alpha}-i\gamma}\right|^2.
\end{equation}
Finally, replacing $\mathcal{R}\rightarrow\mathcal{L}$ and $L\rightarrow0$, we find the expression for the direct current
through the first tunnel junction
\begin{equation}
\langle I_{\mathcal{L}\mathcal{L}}(t)\rangle=\frac{\Gamma |\tau_{\mathcal{L}}|^2}{4\pi^2uv}
\sum_\alpha\left|\int_{-\infty}^0 d\tau \frac{e^{i(\varepsilon_0-i\Gamma)\tau}}{\tau+t+\frac{x_0}{v_\alpha}-i\gamma}\right|^2.
\end{equation}

\section{Derivation of the interference contribution to the current}
\label{Sec:C}
According to the general representation of the current (\ref{small_current_formula}), one needs to find
\begin{equation}
\langle I_{\mathcal{L}\mathcal{R}}(t,t')\rangle=\mathcal{M}(t,t')+\mathcal{M}(t',t),
\end{equation}
where
\begin{align}
\label{M}
\mathcal{M}(t,t')=\langle[\widetilde{A}^{\dagger}_{\mathcal{R}}(t),\widetilde{A}_{\mathcal{L}}(t')]\rangle,
\end{align}
and the average is with respect to the initial state (\ref{Initial_state}).
Applying outlined in the Sec.~\ref{Sec:I} bosonization procedure, we obtain
\begin{align}
\label{A_1}
\langle\widetilde{A}^{\dagger}_{\mathcal{R}}(t)\widetilde{A}&_{\mathcal{L}}(t')\rangle\!=\!\tau_\mathcal{L}\tau_\mathcal{R}^*
\frac{|\tau_d|^2}{a^3}
\!\iint_{-\infty}^0\!\!\!\! d\tau d\tau' e^{i\varepsilon_0(\tau-\tau')}e^{\Gamma(\tau+\tau')} \nonumber \\ \times
\prod_\alpha&\langle0|e^{\frac{i}{\sqrt{2}}\xi_{\alpha u}(x_0+v_\alpha\tau')}
e^{\frac{i}{\sqrt{2}}\xi_{\alpha u}(L-v_\alpha t)} \times \nonumber  \\ &\times
e^{-\frac{i}{\sqrt{2}}\xi_{\alpha u}(-v_\alpha t')}
e^{-\frac{i}{\sqrt{2}}\xi_{\alpha u}(x_0+v_\alpha\tau)}|0\rangle_{\alpha u}
 \nonumber \\ \times&
 \langle0|e^{-\frac{i}{\sqrt{2}}\xi_{\alpha d}(L-v_\alpha t)}
 e^{\frac{i}{\sqrt{2}}\xi_{\alpha d}(-v_\alpha t')}|0\rangle_{\alpha d},
\end{align}
and
\begin{align}
\label{A_2}
\langle\widetilde{A}_\mathcal{L}(t')\widetilde{A}^\dag_{\mathcal{R}}&(t)\rangle\!=\!\tau_\mathcal{L}\tau_\mathcal{R}^*
\frac{|\tau_d|^2}{a^3}
\!\iint_{-\infty}^0\!\!\!\! d\tau d\tau' e^{i\varepsilon_0(\tau-\tau')}e^{\Gamma(\tau+\tau')} \nonumber \\ \times
\prod_\alpha&\langle0|e^{\frac{i}{\sqrt{2}}\xi_{\alpha u}(x_0+v_\alpha\tau')}
e^{-\frac{i}{\sqrt{2}}\xi_{\alpha u}(-v_\alpha t')} \nonumber \\ &\times
e^{\frac{i}{\sqrt{2}}\xi_{\alpha u}(L-v_\alpha t)}
e^{-\frac{i}{\sqrt{2}}\xi_{\alpha u}(x_0+v_\alpha\tau)}|0\rangle_{\alpha u}
  \nonumber \\ \times&
 \langle0|e^{\frac{i}{\sqrt{2}}\xi_{\alpha d}(-v_\alpha t')}
 e^{-\frac{i}{\sqrt{2}}\xi_{\alpha d}(L-v_\alpha t)}|0\rangle_{\alpha d}.
\end{align}

Using the relation similar to Eq.\ (\ref{4_correlator})
\begin{align}\label{4_2_correlator}
&\langle0|e^{\frac{i}{\sqrt{2}}\xi_{\alpha j}(x)}
e^{\frac{i}{\sqrt{2}}\xi_{\alpha j}(y)}e^{-\frac{i}{\sqrt{2}}\xi_{\alpha j}(z)}
e^{-\frac{i}{\sqrt{2}}\xi_{\alpha j}(t)}|0\rangle_{\alpha j}=\nonumber \\ &
\frac{\gamma \sqrt{i(y-x)+\gamma}\sqrt{i(t-z)+\gamma}}{\sqrt{i(z-x)+\gamma}\sqrt{i(t-y)+\gamma}\sqrt{i(z-y)+\gamma}\sqrt{i(t-x)+\gamma}},
\end{align}
and (\ref{2_correlator}), one can easily see that
\begin{equation}
\mathcal{M}(t,t')\sim \delta\left(t'-t+\frac{L}{u}\right)-\delta\left(t'-t+\frac{L}{v}\right).
\end{equation}
Since the integration over $t'$ in the expression (\ref{I_ll}) for the current extends to the region $t'<t$, the term
$\mathcal{M}(t',t)$  does not contribute to the interference current.
Using (\ref{4_2_correlator}) and (\ref{2_correlator}) in Eqs.\ (\ref{A_1}) and (\ref{A_2}) and substituting the result to (\ref{M}), we obtain
\begin{align}
&\langle I_{\mathcal{L}\mathcal{R}}(t,t')\rangle\!=\!\frac{i\Gamma\tau_\mathcal{L}\tau_\mathcal{R}^*}{2\pi^2\sqrt{uv}\eta}
\!\iint_{-\infty}^0\!\!\!\! d\tau d\tau' e^{i\varepsilon_0(\tau-\tau')}e^{\Gamma(\tau+\tau')} \nonumber \\ &\times
\left[\delta\left(t'-t+\frac{L}{u}\right)-
\delta\left(t'-t+\frac{L}{v}\right)\right]\prod_\alpha \mathcal{C}(v_\alpha),
\end{align}
where $\eta=L/v-L/u$, and
\begin{align}
\mathcal{C}(v_\alpha)=&\frac{\sqrt{i(x_0+v_\alpha\tau+v_\alpha t')+\gamma}}{\sqrt{-i(x_0+v_\alpha\tau'+v_\alpha t')+\gamma}}\frac{1}{\sqrt{i(v_\alpha\tau-v_\alpha\tau')+\gamma}} \nonumber \\ \times&
\frac{\sqrt{-i(x_0+v_\alpha\tau'+v_\alpha t-L)+\gamma}}{\sqrt{i(x_0+v_\alpha\tau+v_\alpha t-L)+\gamma}}.
\end{align}
The integration over $t'$ in Eq.\ (\ref{I_ll}) removes the delta-function, and we finally arrive at the expression (\ref{int_I_LR}).

\section{Point splitting procedure}
\label{Sec:D}

To clarify the physical meaning of the point splitting procedure, one can examine the following example.
Let us consider a ground state of a system of strongly-interacting electrons confined in two channels of length $L$,
and describe it in terms of bosonic fields $\phi_\alpha(x)$ with periodic boundary conditions.

Below we investigate the effect of adding $N$ electrons to the first channel, whether the point-splitting procedure is performed or not.
Using the procedure outlined in the main part of the paper [\onlinecite{Levkivskyi_1}], we present the correction to the Wigner function in the first channel as
\begin{align}
\label{W_def}
\Delta W(x,\varepsilon,\delta)=&\frac{1}{4\pi^2}\int dz e^{-i\varepsilon z}
e^{-i\frac{\pi}{L}(\delta+\frac{u+v}{2}z)}\nonumber\\ &\times\frac{e^{i\frac{2\pi N}{L}(\delta+\frac{u+v}{2}z)}-1}{\sqrt{i(\delta+uz)+\gamma}\sqrt{i(\delta+vz)+\gamma}},
\end{align}
where the point splitting parameter $\delta$ is set to 0 in the end of the calculations.
The excess particle density in the first channel can be obtained directly from this
correction, and it takes the natural value $\rho(x)=N/L$.
 The corresponding energy distribution function is
 \begin{equation}
 \label{f_e_vac}
 f(\varepsilon)=L\frac{e^{-i\pi\delta/L}}{2\pi\sqrt{uv}}[e^{2i\delta\Delta\mu/(u+v)}g(\varepsilon-\Delta\mu)-g(\varepsilon)],
 \end{equation}
 where
 \begin{equation}
 g(\varepsilon)=\theta(-\varepsilon)e^{i\frac{u+v}{2uv}\varepsilon\delta}
 J_0\left[\frac{(u-v)\varepsilon\delta}{2uv}\right],
 \end{equation}
 and $\Delta\mu=\pi(u+v)N/L$ is the induced chemical potential shift in the first channel.
 Taking the limit $\delta\rightarrow0$, one obtains
 \begin{equation}
 \label{f_ezero}
 f(\varepsilon)=\frac{L}{2\pi\sqrt{uv}}[\theta(\Delta\mu-\varepsilon)-\theta(-\varepsilon)].
 \end{equation}

 Interestingly, exactly the same result for the energy distribution function can be obtained if one sets $\delta=0$ before
 integration directly in Eq.\ (\ref{W_def}).
 This is because the correction to the energy distribution from finite values of $\delta$ comes from the energies well below the Fermi level. We use this fact in the main part of the paper in order to simplify the calculation of the Wigner function and of the energy distribution function. However, this procedure would lead to a wrong expression for the excess density of the number of particles, $\rho(x)=(u+v)N/2\sqrt{uv}L$, which is one of the manifestations of the well-known anomaly of chiral one-dimensional systems. The only possible solution of this seeming paradox comes from the observation that the remaining part of the particle density is accumulated at the bottom of the conduction band, and results from its shift due to the induced potential at the first channel. This effect cannot be described formally within the bosonization procedure.


\begin{thebibliography}{99}

\bibitem{Halperin} B. I. Halperin, Phys. Rev. B {\bf 25}, 2185 (1982). 

\bibitem{Buttiker} M. B\"{u}ttiker, Phys. Rev. B {\bf 38}, 9375 (1988).  



\bibitem{Bocquillon_1} E. Bocquillon,	V. Freulon,	J-.M Berroir,	P. Degiovanni,	B.~Pla\c{c}ais,	A. Cavanna,	Y. Jin,	and G. F\`{e}ve, Nat. Commun. {\bf 4}, 1839 (2013). 


\bibitem{Neder_1} I. Neder, M. Heiblum, Y. Levinson, D. Mahalu, and V.~Umansky, Phys. Rev. Lett. {\bf 96}, 016804 (2006). 

\bibitem{Litvin_1} L. V. Litvin, H.-P. Tranitz, W. Wegscheider, and C.~Strunk, Phys. Rev. B {\bf 75}, 033315 (2007). 

\bibitem{Preden} Preden Roulleau, F. Portier, D. C. Glattli, P. Roche, A.~Cavanna, G. Faini, U. Gennser, and D. Mailly, Phys. Rev. B {\bf 76}, 161309 (R) (2007). 

\bibitem{Neder_2} I. Neder, F. Marquardt, M. Heiblum, D. Mahalu, and V.~Umansky, Nat. Phys. {\bf 3}, 534 (2007). 

\bibitem{Litvin_2} L. V. Litvin, A. Helzel, H.-P. Tranitz, W. Wegscheider, and C. Strunk,  Phys. Rev. B {\bf 78}, 075303 (2008). 

\bibitem{Bieri} E. Bieri, M. Weiss, O. G\"{o}ktas, M. Hauser, C.~Sch\"{o}nenberger, and S. Oberholzer,  Phys. Rev. B {\bf 79}, 245324 (2009). 

\bibitem{Huynh} P-A. Huynh, F. Portier, H. le Sueur, G. Faini, U. Gennser, D. Mailly, F. Pierre, W. Wegscheider, and P. Roche,
                  Phys. Rev. Lett. {\bf 108}, 256802 (2012). 



\bibitem{Helzel} A. Helzel, L. V. Litvin, I. P. Levkivskyi, E. V. Sukhorukov, W. Wegscheider, and C. Strunk, Phys. Rev. B {\bf 91}, 245419 (2015). 


\bibitem{Degiovanni} P. Degiovanni, Ch. Grenier, G. F\`{e}ve, C. Altimiras, H.~le~Sueur, and F. Pierre, Phys. Rev. B {\bf 81}, 121302 (R) (2010). 

\bibitem{Altimiras_1} C. Altimiras, H. le Sueur, U. Gennser, A. Cavanna, D.~Mailly, and F. Pierre, Nat. Phys. {\bf 6}, 34 (2010). 

\bibitem{Sueur} H. le Sueur, C. Altimiras, U. Gennser, A. Cavanna, D.~Mailly, and F. Pierre, Phys. Rev. Lett. {\bf 105}, 056803 (2010). 

\bibitem{Altimiras_2} C. Altimiras, H. le Sueur, U. Gennser, A. Cavanna, D.~Mailly, and F. Pierre, Phys. Rev. Lett.{\bf  105}, 226804 (2010). 



\bibitem{Tewari} S. Tewari, P. Roulleau, C. Greiner, F. Portier, A. Cavanna, U. Gennser, D. Mally, and P. Roche, arXiv:1503.05057v1 



\bibitem{Levkivskyi_1} I. P. Levkivskyi, E. V. Sukhorukov, Phys. Rev. B {\bf 78}, 045322 (2008). 


\bibitem{Giamarchi} Th. Giamarchi, {\it Quantum Physics in One Dimension} (Oxford University Press, Oxford, 2003). 


\bibitem{Ferraro_1} D. Ferraro, A. Feller, A. Ghibaudo, E. Thibierge, E. Bocquillon, G. F\`{e}ve, Ch. Grenier, and P. Degiovanni, Phys. Rev. B {\bf 88}, 205303 (2013). 

\bibitem{Ferraro_2} D. Ferraro, B. Roussel, C. Cabart, E. Thibierge, G. F\`{e}ve, C. Grenier, P. Degiovanni, Phys. Rev. Lett. {\bf 113}, 166403 (2014).


\bibitem{Henny} M. Henny, S. Oberholzer, C. Strunk, T. Heinzel, K. Ensslin, M. Holland, C. Sch\"{o}nenberger, Science {\bf 284}, 5412 (1999).

\bibitem{Neder_3} I. Neder, N. Ofek, Y. Chung, M. Heiblum, D. Mahalu, and V. Umansky, Nature {\bf 448}, 333 (2007).

\bibitem{Ji} Y. Ji, Y. Chung, D. Sprinzak, M. Heiblum, D. Mahalu, and  H. Shtrikman, Nature {\bf 422}, 415 (2003).

\bibitem{Roulleau_1} P. Roulleau, F. Portier, P. Roche, A. Cavanna, G. Faini, U. Gennser, and D. Mailly, Phys. Rev. Lett. {\bf 100}, 126802 (2008). 

\bibitem{Roulleau_2} P. Roulleau, F. Portier, and P. Roche, A. Cavanna, G.~Faini, U. Gennser, and D. Mailly, Phys. Rev. Lett. {\bf 102}, 236802 (2009). 

\bibitem{Bocquillon_3} Erwann Bocquillon, Vincent Freulon, Fran\c{c}ois D. Parmentier, Jean-Marc Berroir, Bernard Pla\c{c}ais, Claire~Wahl, J\'{e}rome Rech, Thibaut Jonckheere, Thierry Martin, Charles Grenier, Dario Ferraro, Pascal Degiovanni, and Gwendal F\`{e}ve, Ann. Phys. (Berlin) {\bf 526}, 1-2,1-30 (2014). 

\bibitem{Geerligs} L. J. Geerligs, V. F. Anderegg, P. A. M. Holweg, J.~E.~Mooij, H. Pothier, D. Esteve, C. Urbina, and M.~H.~Devoret, Phys. Rev. Lett. {\bf 64}, 2691 (1990). 

\bibitem{Pothier} H. Pothier, P. Lafarge, C. Urbina, D. Esteve, and M.~H.~Devoret, Europhys. Lett. {\bf 17}, 249 (1992). 


\bibitem{Kouwenhoven} L. P. Kouwenhoven, A. T. Johnson, N. C. van der Vaart, C. J. P. M. Harmans, and C. T. Foxon Phys. Rev. Lett. {\bf 67}, 1626 (1991). 


\bibitem{Fujiwara} A. Fujiwara, N. M. Zimmerman, Y. Ono, and Y. Takahashi, Appl. Phys. Lett. {\bf 84}, 1323 (2004). 


\bibitem{Feve} G. F\`{e}ve, A. Mah\'{e}, J.-M. Berroir, T. Kontos, B. Pla\c{c}ais, D. C. Glatti, A. Cavanna, B. Etienne, and Y. Jin, Science
{\bf 316}, 1169 (2007).

\bibitem{Bocquillon_2} E. Bocquillon, V. Freulon, J.-M. Berroir, P. Degiovanni, B. Pla\c{c}ais, A. Cavanna, Y. Jin, and G. F\`{e}ve, Science {\bf 339}, 1054 (2013).  

\bibitem{Levkivskyi_2} I. P. Levkivskyi, E. V. Sukhorukov, Phys. Rev. B {\bf 85}, 075309 (2012). 

\bibitem{Levkivskyi_3} I. P. Levkivskyi, E. V. Sukhorukov, Phys. Rev. B {\bf 109}, 246806 (2012). 

\bibitem{Jullien} T. Jullien, P. Roulleau, B. Roche, A. Cavanna, Y. Jin, and D. C. Glattli, Nat. Phys. {\bf 514}, 603–607 (2014).

\end{thebibliography}
\end{document}